\documentstyle[12pt]{article}
\begin{document}
\newcommand{\be}{\begin{equation}}
\newcommand{\ee}{\end{equation}}
\newcommand{\ba}{\begin{eqnarray}}
\newcommand{\ea}{\end{eqnarray}}
\vskip 5.5cm
\centerline{\Large \bf The inequivalence of thermodynamic
ensembles}
\vskip 2.5cm
\footnotetext[1]{Present address: Laboratoire de Physique Theorique de l' Ecole
Normale Superieure, 24 Rue Lhomond, 75005 Paris, France (E-mail:
parenta@physique.ens.fr)}
\centerline{\large Renaud Parentani\footnotemark[1]}
\vskip 1.5cm
\centerline{The Racah Institute of Physics, The Hebrew
University of Jerusalem,}
\centerline{Jerusalem 91904, ISRAEL}
\bigskip
\vskip 4.5cm
\centerline{\large\bf Abstract}
\bigskip
The inequivalence of thermodynamical ensembles related by
a Legendre  transformation is manifest in self-gravitating
systems and in black hole  thermodynamics. Using the
Poincar\'e's method of the linear series, we describe  the
mathematical reasons which lead to this inequivalence
which in turn induces a hierarchy of ensembles: the
most stable ensemble describes the  most isolated system.
Moreover, we prove that one can obtain the degree of
stability of all equilibrium configurations in any
ensemble related by Legendre transformations to the most
stable if one knows the degree of  stability in the most
stable ensemble.

\newpage
\section*{1. Introduction}

\hspace{\parindent}
In self-gravitating systems$^{[1-5]}$ or in systems with
highly degenerate spectra such as strings$^{[6]}$ or black
holes$^{[7-14]}$, it is well known that thermodynamical
ensembles are generally inequivalent. This means
that  two ensembles related by a single Legendre
transformation (or a Laplace transformation in
Statistical Mechanics) become unstable in different
situations. For instance there exist stable situations in
microcanonical ensembles with negative specific heat.
Surprisingly there also exist situations in  which
canonical ensembles with positive specific heat are
unstable.

Thus, evidently, the heat capacity alone does not control
the stability of those ensembles and one needs more
reliable criteria.  These can be provided,  under specific
conditions (see refs.~1,~2), by the Poincar\'e's linear
series method.  This method offers the great advantage to
work without having to solve the  eigenvalue equation which
control the quadratic fluctuations ({\it i.e.} without having
to
calculate  the Poincar\'e's coefficients of stability).
The method provides also, as we will show, a useful guide
to obtain the stability limits of any Legendre transformed
ensemble when one knows the stability range of one
ensemble and this,  whether or not these ensembles are
equivalent.

In this article we will study the origin as well as the
generic properties  of this inequivalence. Addressing the
determination of stability criteria, we  will first find
the mathematical reason which leads to the breaking of
the  apparent ``symmetry" between an ensemble and its
Legendre transform. This  is strange indeed in that if we
apply a second Legendre transformation one is led back
to the initial ensemble. This mathematical reason will
also permit us  to make contact with physics because the
most stable ensemble will  systematically describe the
most isolated system. By isolated we mean an  ensemble
whose variable (called in this article control parameter)
which characterizes it, is a conserved quantity of the
hamiltonian of the system. On the  contrary, non-isolated
systems are characterized by control parameters kept  fixed
by the intervention of an external reservoir.  Familiar
examples of  couples of control parameters are
respectively: energy-temperature, particle
number-chemical potential, volume-pressure, angular
momentum-angular velocity.
\newpage
We then demonstrate the following
points:

\begin{enumerate}
\item The inequivalence comes from the fact that the less
stable ensemble has a  richer spectrum of fluctuations
which contains {\it one} extra fluctuating quantity  (or
one more Poincar\'e stability coefficient).\

\item From this fact one immediately concludes that when
the corresponding  eigenvalue of those extra fluctuations
is positive it simply means that the  number of unstable
modes in the two ensembles is identical (This number can
be zero in which case both ensembles are stable).\

\item More remarkable is the fact that it is always this
new eigenvalue which  encodes the changes of stability of
the less stable ensemble. This means that  whenever the
more stable system approaches instability ({\it i.e.} when
its lowest  eigenvalue tends to zero) an algebraic identity
dictates that the new eigenvalue  will cross zero before
the lowest one of the more stable system.\
\end{enumerate}

Our analysis results in a qualification of the traditional
criterion wherein the positivity of the specific heat was
considered as a sufficient condition for the stability of
the canonical ensemble. It provides also a timesaving
procedure when, being aware that ensembles might be
inequivalent, one wishes to determine stability ranges
under various conditions$^{[4, 5, 14, 15]}$. As an
illustration of its power, we apply our analysis to the
situation (recently studied by Kaburaki {\it et
al.}$^{[10, 11]}$) in which the angular  momentum of
rotating black holes has a stabilizing effect. We derive
results on stability that are stronger than those
previously published.

\section*{Section 2}

\hspace{\parindent}
Our goal is to compare the stability limits of two
ensembles related by a  single Legendre transformation.
The demonstration of the three points  mentioned in the
Introduction proceeds as follows. We start by the analysis
of  a simplified mathematical abstraction in which all the
members of the  first ensemble are states parametrized by a
single variable, denoted by $x$, at a fixed  control
parameter $s$.
Subsequently the applicability of the
mathematical  exercise to physical situations including
many degrees of freedom (as in  thermodynamics) is
discussed.

In the first ensemble,
equilibrium situations and stability conditions are
provided by a  potential function $F(x;s)
$\footnote{For the reader who wants to have in mind a specific example,
$s$ may be interpreted as the total energy $E$ (then by working
at fixed $s$ one has the microcanonical ensemble), the variable
$x$ viewed as the energy repartition between two subsystems and
the potential $F(x;s)$ may be interpreted as the entropy out of equilibrium
when the variable $x$ is not at its equilibrium value $X_i (s)$.
Then $t$ is the inverse temperature and the potential $-G$ is the free
energy divided by the temperature. The reader might usefully consult ref. 19
to see in detail how this method apply to situations containing a black hole
surrounded by radiation.}.
The Legendre
transformation to a new potential  $G(x;t)$ with control
parameter $t$, is given by:

\be
G(x;t) = F(x; S(x;t)) - tS(x;t)
\ee

\noindent
where $S(x;t)$ is the solution (unique at fixed $x$) of:

\be
t = \partial _sF(x;s)
\ee

\noindent
$\partial$ designates partial derivative with all other
quantities kept fixed.
The equilibrium configurations,
$X_i(s) \ \ i=1, \cdots n,$ are the solutions of:

\be
\partial_xF(x;s)=0
\ee

\noindent
Those linear series of equilibrium configurations coincide
with the solutions  of

\be
\partial_xG(x;t)=0
\ee

\noindent
by virtue of eq(2). At this point, the equivalence of the
two ensembles is manifest since $F$ is the Legendre
transform potential of $G$, with the role of $s$ and $t$
interchanged, and since they determine the same
equilibrium  configurations. (We shall  adopt the notation
$F$-ensemble when the description of the states is given
by $F(x;s)$ and $G$-ensemble for the other).

A dissymmetry between the two is introduced by formulating
a  necessary and sufficient condition for the
{\it stability} of equilibrium  configurations of one
ensemble, say the $F$-ensemble:

\be
\partial_x^2F(x=X_i(s);s)<0
\ee

\noindent
({\it i.e.} that $F$ be a maximum). Thus any change of
stability, in the $F$-ensemble,  will occur if and only
if  $\partial ^2_xF$ crosses zero.

Contrariwise, in the $G$-ensemble, at fixed $t, s$
fluctuates\footnote{We are using the thermodynamical terminology
even though the present analysis applies as well to mechanical stability,
thus by 'fluctuates' we mean that we have to consider neighboring
solutions around the equilibrium one in ordre to determine stability.
In the same spirit by 'fluctuations of $x$' we shall designate the quantity
$(-\partial_x^2F)^{-1/2}$ which is the RMS fluctuations
of $x$ in a statistical
ensemble weighted by $e^{F}$. We refer again to ref. 19 for explicit examples
in which the notions of stability and fluctuations are defined more
concreately.}
as well around  the solution of eq(2):
$S(x,t)$. Thus further conditions must be met to fulfill
stability ({\it i.e.} $G$ has to be a true maximum and not
only a saddle point in order to insure that the new
fluctuations be bounded). Since those new conditions  need
not be met even when eq(5) is satisfied, the $G$-ensemble
is the less stable one. Our purpose is now to prove that
those new conditions inevitably restrict the range of
stability of equilibrium situations in the $G$-ensemble.
(At the end of the paper, we shall discuss the ``opposite"
problem, that is, how, when one's consideration begins
with the knowledge of the stability limits of the less
stable  ensemble, one may compute the extension of the
stability range for the more  stable ensemble.)
\bigskip

We will first review how changes of stability, in the
$F$-ensemble, manifest themselves using Poincar\'e
method. The reader unfamiliar with this method may usefully consult
refs [1,2,19]. In addition, we are using the same notations and conventions.
 The reason for this review is that it affords an
instructive illustrative example of how the topological
behavior of the linear series in the vicinity of a
critical point {\it inevitably} restricts the stability
range of the Legendre transformed ensemble.

The idea of the method is to construct the linear series,
locus of  solutions of eq(3), as a function of $s:
x=X_i(s)$. Poincar\'e theorem states that  stability
changes may occur only at a bifurcation ({\it i.e.} when
two series intersect) or at a turning point ({\it i.e.}
when two series merge into each other). Bifurcations
occur because of an ``excess" of symmetry and can be
removed by the slightest  modification of the potential
$F(x;s)^{[2,16]}$. Hence, only the behavior
of the series near a turning point is studied here.

More specifically we consider the particular
situation where one has
two  equilibrium solutions of eq(3) for $s$ smaller than a
certain maximum denoted $s_f$. These solutions merge into
each other as $s$ tends to $s_f$ and do not exist for $s$
greater than $s_f$. (This situation corresponds to the
case $i$ (Fig. 2) of the  classification of ref.1) The
series of solutions $x=X_1(s)$ describes stable
equilibrium situations for which eq(5) is satisfied, and
the series $x=X_2(s)$ describes unstable situations. The
method then proceeds by evaluating the second
total derivative along those series:

\be
{{d^2F_a}\over{ds^2}} = \partial ^2_sF_a + \partial_x
\partial _s F_a {{dX_a}\over{ds}}
\ee

\noindent
where \( F_a \equiv F(X_a(s),s)\) and \(a = 1,2\).
With the help of the following identity

\be
{{d}\over{ds}}\left(\partial_xF_a\right) = 0 =
\partial_s\partial_xF_a + \partial_x^2F_a
{{dX_a}\over{ds}}
\ee
eq(6) becomes

\be
{{d^2F_a}\over{ds^2}}=
\partial^2_sF_a -
\left(\partial_x\partial_sF_a\right)^2/\partial^2_xF_a
\ee

\noindent
Hence  the change of stability at \(s=s_f\),
where \(\partial^2_xF_a=0\), is manifested by the
divergence of this second derivative unless

\be
\partial_x\partial_sF_a(s=s_f)=0
\ee

\noindent
(We do not consider this degenerate possibility,
which leads to a bifurcation, since it
has been proven$^{[2,16]}$ that the slightest modification of
\(F\) will destroy the simultaneous vanishing of
\(\partial^2_xF_a\) and \(\partial_s\partial_xF_a\)). We
emphasize the role and the physical meaning of eq(7).
If \(\partial_x\partial_sF_a \not=0\), one has

\be
{{dX_a}\over{ds}} \sim \left( \partial^2_xF_a\right) ^{-1}
\ee

\noindent
which means that near the instability, the evolution of
the {\it equilibrium} value \(X_a(s)\) (or the mean value
in thermodynamical ensembles) is entirely controlled by
the growing of the {\it fluctuations}. Hence, the
knowledge of  \(X_a(s)\) is sufficient
to determine the stability changes without
computing separately the fluctuations themselves.

Furthermore since

\be
{{d^2F_a}\over{ds^2}}
\mathrel{\mathop{\kern0pt\longrightarrow}\limits_{s\to
s_f}} (-) \left(\partial^2_xF_a\right)^{-1} \quad \qquad
{{\to +\infty~{\rm along}~ X_1(s)}\atop{\to -\infty~{\rm
along}~X_2(s)}}
\ee

\noindent
the linear series 1, describing stable situations, lies
inevitably below (in the $s-t$ plane) the series 2,  as depicted in Fig.1.

Before examining what are the consequences of this unique
behavior,  we briefly discuss what could have happened in
more complex situations.\\
\begin{enumerate}
\item For ensembles with many \(x\) variables, the linear
series 1 could  describe already unstable situations with
\(n\) negative modes (where \(n<\) number of degrees of
freedom described by the \(x\) variables). Nevertheless by
going anti clock wise along the curve of Fig.1. one still
increases the number of unstable modes each time
\(\partial_sF_a\) has a vertical tangent.

\item In the cases where one has many parameters \(s\),
each of them or any  non singular function of them can be
used to study the changes of stability of the
\(F\)-ensemble, leading to identical predictions because
the vertical tangent  only occurs when
\(\partial^2_xF\) vanishes. But if one wants to study the
stability of a  particular Legendre transformed ensemble
and compare it to the stability of  the initial ensemble,
it is appropriate to select the parameter which is used
to  define the Legendre transformation. (In the present
case with only one  parameter, we are automatically in
such a situation.)
\end{enumerate}

We now examine the stability in the \(G\)-ensemble. We
first redefine in a  slightly different manner the Legendre
transformation, eq(1), to see clearly  how the
transformation from \(F\) to \(G\) enlarges the spectrum of
fluctuations by one eigenvalue\footnote{We recall that in a canonical
ensemble the total energy $E$ (here $s$) of the system fluctuates
due to the contact with the heat-reservoir.} (in our case we have two
fluctuating quantities \(x\) and \(s\)) and  why the new
fluctuations always encode the changes of stability of the
\(G\)-ensemble.

\be
G(x,s;t)=F(x;s)-st
\ee

\noindent
Equilibrium configurations are now provided by

\ba
\partial_xG&=&0=\partial_xF \\
\partial_sG&=&0=\partial_sF - t
\ea

\noindent
Eq(13) leads identically to eq(3) since now the only
\(x\) dependence of \(G\) is through \(F\).  Eq(14)
connects \(s\) with its Legendre-conjugate \(t\) as in eq(2).
These two equations furnish: \(x=X_a(t) =
X_a(S_a(t))\), where \(S_a(t)\) are the two values
of the solution of eq(14)
evaluated at the equilibrium situations $x=X_a$.

The range of stability is obtained as previously by
computing the second  derivative
 of \(G\), with respect to $t$,
along the linear
series \(x=X_a(t)\), see eqs(6-11)

\be
{{d^2G_a}\over{dt^2}} = \partial^2_tG_a +
\partial_x\partial_tG_a {{dX_a}\over{dt}} +
\partial_s\partial_tG_a{{dS_a}\over{dt}}
\ee

\noindent
where \(G_a \equiv G(X_a(t),S_a(t);t)\). Since by
construction in eq(12) one has

\be
\partial^2_tG=0~,\quad \partial_t\partial_sG=0~, \quad
\partial_s\partial_tG=-1
\ee

\noindent
one finds

\ba
{{d^2G_a}\over{dt^2}}&=& -{{dS_a}\over{dt}}\nonumber\\
&=&-\left(\partial^2_sF_a -
(\partial_x\partial_sF_a)^2/\partial^2_xF_a\right)^{-1}\\
&=&-\left({{d^2F_a}\over{ds^2}}\right)^{-1}=
-\left({{dT_a}\over{ds}}\right)^{-1} \nonumber
\ea

\noindent
where \(T_a(s)=\partial_sF_a\) and where the second
equality follows, as for the \(F\)-ensemble (see eq(7)),
from

\ba
&&{{d}\over{dt}}\left(\partial_xG_a\right)=0=
\partial^2_xF_a{{dX_a}\over{dt}}+\partial_s\partial_xF_a
{{dS_a}\over{dt}}\nonumber\\
&&{{d}\over{dt}}\left(\partial_sG_a\right)=0=-1+
\partial_x\partial_sF_a{{dX_a}\over{dt}}+
\partial^2_sF_a{{dS_a}\over{dt}}
\ea

\noindent
We thereby recover the well-known identity,
eq(17), between first total derivatives. The reasons
that we prove this identity are the following:

\begin{enumerate}
\item By the very structure of the Legendre transformation
which implies  eqs(16), we see that the fluctuations of
\(s\) ( {\it i.e.} $(- d^2 F/ds^2)^{-1/2}$ )
alone control the vertical tangents of
\(\partial_tG_a=S_a(t)\). But they also control, as
we will prove, the changes of stability  in the
\(G\)-ensemble.

\item These new fluctuations encode in a very specific way
the previous fluctuations of \(x\) (see eq(17)) and
one has still to impose eq(5) in order to  have stable
equilibrium (even though, there will be no new unstable
mode, in the \(G\)-ensemble, when \(\partial^2_xF_a\) will
become negative, see the Appendix).
\end{enumerate}

{}From eq(17) we obtain that changes of stability in the
\(G\)-ensemble may occur when \(d^2F_a/d^2s=0\) and that
(since only maxima describe stable equilibrium) the more
stable series is the one with \(d^2F_a/d^2s<0\), {\it
i.e.} see eq(8) when:
\be
\partial_s^2F_a <
\left(\partial_x\partial_sF_a\right)^2/\partial^2_xF_a
\ee

\noindent
Since \(\partial^2_xF_a\) is negative along the stable
linear series 1 and tends to \(-\infty\) as \(s\)  tends
to \(s_f\), the inequality (19) has to break down {\it
before} \(s_f\) at some maximum point \(s=s_g\). (We
assume that \(F\) has regular second derivatives). Hence,
since \(d^2F_a/ds^2\) crosses zero before
\(\partial^2_xF_a\), the new fluctuations will {\it
always} control the changes of stability of \(G\). This
proves the third and the  last point mentioned in the
introduction.

One can visualize this ordering by examining the behavior
of the linear series in the vicinity of \(s_f\) and
\(s_g\). Only two situations may occur. Either eq(19) is
never satisfied and the \(G\)-ensemble is always unstable
with one negative mode (case I of Fig.2.  An example of
this case is provided by a Black Hole in contact with a
heat bath.$^{[7]}$), or eq(19) is satisfied until
\(s=s_g\) and the \(G\)-ensemble is stable up to that point
(case II, see the example of star cluster in ref. 1).

For the interested reader we present in the appendix the
(more traditional) analysis of the stability of the
\(G\)-ensemble when any reference to \(s\) has been
eliminated, through the use of eq(2), before looking at
equilibrium. The result one obtains is that the quadratic
fluctuations of \(x\) now evaluated at fixed \(t\), rather
than fixed \(s\), encode automatically the determinant
factor \(d^2F_a/ds^2\) hence leading again to the same
range of stability. In addition we present in this
appendix how, starting from \(G\), one recovers the correct
enlarged range of stability for the \(F\)-ensemble by an
inverse Legendre transformation. The mathematical reason
for which one {\it increases} the stability range, by
going back from \(G\) to \(F\), is directly related to the
fact that when both ensemble are stable one has

\be
{{d^2F_a}\over{ds^2}} < 0~, \quad {{d^2G_a}\over{dt^2}} =
-\left({{d^2F_a}\over{ds^2}}\right)^{-1} >0
\ee

\noindent
This inevitable discrepancy of signs has the following
physical interpretation.  Parameters of the \(s\) type,
like energy, angular momentum or volume represent
conserved quantities of the Hamiltonian of an isolated
system and have to be treated as the variables \(x\)
(stability upon extremization, eq(14), requires the same
sign for the curvature, see eq(19)). In contrast
parameters of the \(t\) type, like temperature, angular
velocity or pressure, are kept fixed by the  intervention
of an external ``reservoir". Furthermore they enter only in
a  linear way into the state function (see eq(12)) and
play an identical role to Lagrange multipliers. The
fact that their extrema correspond to minima, eq(20),
translates in Statistical Mechanics into the fact
that inverse  Laplace transformations behave like Fourier
transformations\(^{[15]}\). Thus their integration, due to their
linear dependence, leads to a Dirac \(\delta\)-function
which reduces by {\it one} the fluctuating
quantities.

In thermodynamics, in complex situations, one deals with
many  control parameters. The ensemble with all its
parameters being of the \(s\) type is  the most stable one
and describes the most isolated system. The stability of
this  ensemble does not refer to any other ensemble and the
extremization of its  state function is a rephrasing of
the Second Principle of Thermodynamics\footnote{This was
pointed out to me by R. Brout.}. Contrariwise the
stability of all  other ensembles is subject not only to
additional necessary conditions (like the  positivity of
the specific heat, see eq(19)) but also to the stability
of the most stable ensemble, eq(5).\\

{\it Application to rotating black holes}\\
To illustrate the power and the simplicity of our analysis
we consider the
situation in which one has a rotating black hole in
contact with a heat bath. Working at fixed  angular
momentum, it has been noted that there exists a temperature
at which the specific heat changes sign\(^{[8,9,10]}\). The
question is then: is the equilibrium, in the canonical
ensemble, stable or only less unstable when the specific
heat is positive?

The problem of the stability of a rotating black hole was
addressed in  ref(10). In the microcanonical ensemble
({\it i.e.} at fixed mass \(M\), fixed angular momentum
\(J\) and with the entropy \(S\) as the relevant state
function), it was  stated that the absence of vertical
tangent in the plot \(\beta=\partial_MS\) versus \(M\)
indicates that their is no change of stability when one
varies the ratio \(J/M^2\) (see Fig. 3). Then, since it
had been proven that isolated Schwarzschild black holes
are stable\(^{[17]}\), it was correctly concluded that all
isolated rotating black holes are stable\footnote{Nevertheless,
since the entropy \(S(M,J)\) is already an extremized
quantity  with respect to the unknown variables ``\(x\)"
which describe the microscopical states of the black
hole, one has to {\it assume}
 that a simultaneous vanishing
(see eq(9) and the associated discussion) which could hide
changes of stability does not occur.}.

In the canonical ensemble, by virtue of this analysis and
our statement 2 (see Introduction), one immediately
concludes that the flip of sign of the specific heat
at \(\beta = \beta_c\) (see Fig. 3) does correspond to a
real change of stability and that all equilibrium
situations along the linear series with the smaller mass
are stable. This conclusion is stronger that the one given
in ref.(10) because the authors did not take into account
the fact that the instability of the Schwarzschild black
hole in contact with a heat bath is due to a {\it single}
instable mode.

\appendix
\renewcommand{\thesection}{Appendix
\Alph{section}}
\section{}

Two points will be discussed in this appendix. The first
one is the  recovering of the changes of stability of the
\(G\)-ensemble when all dependence of \(s\) has been
eliminated prior looking at equilibrium configurations.
The  second point concerns the extension of the stability
range when, starting from \(G\), one wishes to analyze the
stability of the \(F\)-ensemble.\\

Since all explicit dependence of \(s\) has been eliminated,
one has to analyze the stability of \(x\) fluctuations at
fixed \(t\) rather than fixed \(s\). We have  nevertheless
to verify that the fluctuations of \(s\) at fixed \(x\)
described by \(\partial^2_sF\) are bound ({\it i.e.}
\(\partial^2_sF<0\). We assume here that this is
satisfied). The state function which control the \(x\)
fluctuations is (see eq(1)):

\renewcommand{\theequation}{\Alph{section}.
\arabic{equation}}
\setcounter{equation}{0}

\be
G(x;t) = F(x;S(s;t)) - tS(x;t)
\ee

\noindent
where \(S(x;t)\) is the solution of eq(2). Equilibrium
situations are furnished (see eqs(3,4)) by:

\be
\partial_xG|_t=0=\partial_xF|_s
\ee

\noindent
by virtue of eq(2). (In this appendix, in order to prevent
any quibble we specify by a vertical line which quantity
is kept fixed upon derivation.)

Stable equilibrium requires \(G\) to be a maximum (see
eqs(5,17,19))

\be
\partial^2_xG_a|_t < 0
\ee

\noindent
This condition reads

\be
\partial^2_xG_a|_t = \partial^2_xF_a|_s +
\partial_s(\partial_xF_a|_s)|_x \left(dS_a/dx\right)|_t
\ee

\noindent
where \(S_a(t)=S(X_a(t),t)\). $\left(dS_a/dx\right)|_t)$ can be obtained
by taking the total differential of eq(2):

\be
\partial_x(\partial_sF|_x)|_s dx + \partial^2_sF|_x ds
= dt = 0
\ee

\noindent
hence

\[
{{dS}\over{dx}}|_t = -
(\partial_x\partial_sF)/(\partial^2_sF)
\]

\noindent
Then eq(A.4) becomes

\ba
\partial^2_xG_a&=&\partial^2_xF_a -
(\partial_x\partial_sF_a)^2/\partial^2_sF_a\nonumber\\
&=&\partial^2_xF_a
\left({{d^2F_a}\over{ds^2}}\right)(\partial^2_sF_a)^{-1}
\ea

\noindent
where we have used eq(8). By virtue of the analysis which follows eq(19)
and which indicates that \(d^2F_a/ds^2\) crosses always
zero before \(\partial^2_xF_a\), one recovers that the
changes of stability of the \(G\)-ensemble occur
when \(d^2F_a/ds^2\) vanishes. Furthermore, eq(A.6)
indicates that the vanishing of \(\partial^2_xF_a\) at
\(s=s_f\) does not lead to the vanishing of
\(\partial^2_xG_a\) because of the divergence
of \(d^2F_a/ds^2\) at that point (see  eq(11)).
Finally, by an analysis similar to the one given after
eq(18) one can  easily show that the {\it second} negative
eigenvalue of the \(G\)-ensemble will always appear after
the first one of the \(F\)-ensemble, thereby reinforcing
the point 1 (see Introduction) by introducing a well
defined ordering.\\

The second point of this appendix concerns the following
problem:  having obtained the limit of the stable
configurations for the less stable ensemble (when A.6 vanishes)
how can one
evaluate the extended range of stable configurations in
the \(F\)-ensemble? This problem is by no means purely
academical: Under certain circumstances the evaluation
of quadratic fluctuations can be performed easily only in,
say, the grand canonical ensemble. Then two questions
arise when one wants to use those results into the
evaluation of the stability limits for more stable
ensembles (see refs. [4,5,15], where this problem  was
encountered and discussed). The first one concerns the
possibility of extending the evaluation of the
fluctuations outside the stability domain of the ensemble
in which they were computed. The second question is: how the
instability  of the initial ensemble will manifest itself
into the expression of the stability of the fluctuations
in a more stable ensemble?  Answering those questions can
be achieved in many different ways. We only sketch two
possible ways using our simple mathematical abstraction.

The most pedestrian way takes the lines of the analysis
presented from  eq(12) to eq(19): one introduces a third
fluctuating quantity ({\it i.e.} \(t\)) and one finds  that
the role of its fluctuations is to suppress completely the
fluctuations of \(s\)  thereby leading back to the
``initial" \(F\)-ensemble with its single fluctuating
variable: \(x\). (This procedure corresponds to the second
(noted II)  diagonalization scheme of ref. [15]).

A second illuminating method proceeds, as shown above (see
eq(A.6)), by expressing the quadratic fluctuations of
\(x\) at fixed \(s\) in terms of the fluctuations
evaluated at fixed \(t\). One has:

\be
\partial^2_xF_a=
\partial^2_xG_a\left({{d^2G_a}\over{dt^2}}\right)
(\partial^2_tG_a)^{-1}
\ee

\noindent
Suppose now that one increases \(s\), starting from stable
configurations for \(G\) ({\it i.e.} for \(s<s_g\), see
Fig. 2, where \(\partial^2_xF_a<0\)). When \(s\) crosses
\(s_g\) both \(\partial^2_xG_a\) and \(d^2G_a/dt^2\) flip
sign but their product is perfectly well defined. Hence
\(\partial^2_xF_a\) remains negative and
will vanish only when \(d^2G_a/dt^2\)
vanishes. Eq(A.7) proves that the evaluation of the \(x\)
fluctuations in the \(F\)-ensemble (controlled by
\(\partial^2_xF_a\)) can be computed safely in the
\(G\)-ensemble even in situations which correspond to
unstable equilibrium in \(G\). (This simultaneous
vanishing, at \(s=s_g\), is precisely what was encountered
in ref. 4, and subsequently  analyzed in refs. 5 and 15. It
leads to the first (I) diagonalization scheme of ref. 15.)

\bigskip
\noindent
{\bf Acknowledgements}

We would like to thank J. Katz for his careful
explanations of the Poincar\'e's method and for many
useful remarks. We also thank R. Brout for interesting
discussions about the physical content of the conclusions
reached in this article.

\newpage
\renewcommand{\thesection}
\section{\Large\bf References}
\newcounter{refer}
\begin{list}{$^{\arabic{refer}}$}{\usecounter{refer}}
\item J. Katz, Mon. Not. R. Soc. Lond. {\bf 183} (1978)
765.
\item J. Katz, Mon. Not. R. Soc. Lond. {\bf 189} (1979)
817.
\item D. Lynden-Bell and R. Wood, Mon. Not. R. Soc.
Lond. {\bf 138} (1968) 495.
\item G. Horwitz and J. Katz, Astroph. J. {\bf 211} (1977)
226.
\item G. Horwitz and J. Katz, Astroph. J. {\bf 222} (1978)
941.
\item Y. Aharonov, F. Englert and J. Orloff, Phys. Lett.
{\bf B199} (1987) 366.
\item S. Hawking, Phys. Rev. {\bf D13} (1976) 191.
\item P. Davies, Proc. R. Soc. Lond. {\bf A353} (1977)
499.
\item B. Schumaker, W. Miller and W. Zurek, Phys. Rev.
{\bf D46} (1992) 1416.
\item O. Kaburaki, I. Okamoto and J. Katz, Phys. Rev.
{\bf D47} (1992) 2234.
\item J. Katz, I. Okamoto and O. Kaburaki, Class. Quant.
Grav. {\bf 10} (1993) 1323.
\item J. York, Phys. Rev. {\bf D33} (1986) 2092.
\item R. Balbinot and A. Barletta, Class. Quant. Grav.
{\bf 6} (1989) 195, 203.
\item G. Comer, Class. Quant. Grav. {\bf 9} (1992) 947.
\item G. Horwitz, Commun. Math. Phys. {\bf 89} (1983) 117.
\item J. M. T. Tompson, Phil. Trans. R. Soc. Lond. {\bf 292}
(1979) 1386.
\item C. Misner, K. Thorne and J. Wheeler,
Gravitation. Freeman (1973).
\item H. Callan, Thermodynamics. Wiley International
(1960).
\item R. Parentani, J. Katz and I. Okamoto, Thermodynamics of a black hole in a
cavity
 (1994) gr-qc 9410015

\end{list}

\newpage
\centerline{\bf Figure captions}

\bigskip
\noindent
Fig. 1. Plot of the equilibrium curve \(T_a(s) \equiv
\partial_sF_a(s)\), evaluated along the linear series
\(X_1(s)\) and \(X_2(s)\), in the vicinity of the turning
point \(s=s_g\). The series \(X_1(s)\) describe stable
equilibrium configurations and lie inevitably under the
series \(X_2(s)\). At \(s=s_g\), the fluctuations of \(x\)
are unbound (\(\partial_x^2F_a\) vanishes) and this
manifests itself by a vertical tangent of \(T_a(s)\). This
can be understood by  extending the definition of the
parameter \(t\) away from equilibrium: \(t(x,s) \equiv
\partial_sF\), and letting it fluctuate around \(T_a(s)\)
according to the \(x\) fluctuations$^{[19]}$.\\

\noindent
Fig. 2. This is the same plot as the one of Fig. 1. The
two possible  behaviors of \(T_a(s)\) are displayed. Along
the linear series (1,I) the \(G\)-ensemble  is always
unstable with one negative eigenvalue. Along (1,II) the
\(G\)-ensemble is stable for \(s<s_g\) (or \(t<t_g\)).
The horizontal tangent of \(dS_a/dt\) at \(t=t_g\),
indicates that the fluctuations of the variable \(s\) are
infinite. This should be contrasted with the parametrical
fluctuations of \(t\) in the \(F\)-ensemble.\\

\noindent
Fig.3. Plot of the equilibrium curve \(\beta
= \partial_MS(J,M)\) versus the mass \(M\), at
fixed  angular momentum \(J\) and describing a Kerr black
hole$^{[7,9]}$. At \(\beta=\beta_c\), \(\partial_M\beta\)
flips sign. Below \(M_c\), in the canonical ensemble, the
black hole is in  stable equilibrium with the heat bath.
This situation corresponds to the case II  of Fig. 2, with
\(s_f\)  rejected at infinity.

\end{document}